\documentclass[aps,pre,amsmath,amsfonts,amssymb,superscriptaddress,showpacs,11pt]{revtex4}
\usepackage{graphics}
\usepackage[latin1]{inputenc}
\usepackage{amscd}
\usepackage{soul}

\usepackage[pdftex]{graphicx}

\usepackage{rotating}
\usepackage[pdftex,hypertexnames=true,linktocpage=true]{hyperref}

\usepackage{color}

\usepackage{amsthm}
\usepackage{framed}
\usepackage{amssymb}
\usepackage{amsmath,mathrsfs}
\usepackage{hhline}
\usepackage{ulem}
\usepackage{fancyhdr}
\usepackage{tikz}
\everymath{\displaystyle}
\everymath{\displaystyle}

\everymath{\displaystyle}
\newcommand{\beq}{\begin{equation}}
\newcommand{\eeq}{\end{equation}}

\newcommand{\bi}{\begin{itemize}}
\newcommand{\ei}{\end{itemize}}
\newcommand{\D}{\mathrm{D}}

\newcommand{\KL}{\mathrm{KL}}
\newcommand{\Q}{\mathrm{Q}}
\newcommand{\Hilb}{\mathcal{H}}

\def\RR{\mathbb{R}}

\def\NN{\mathbb{N}}

\newcommand{\Tau}{\mathcal{T}}

\newcommand{\Ma}{\mathrm{M}}
\newcommand{\esse}{\mathcal{S}}

\newcommand{\Div}{\mathcal{D}}
\newcommand{\nihat}{\mathrm{X}}

\newcommand{\tangent}{{\rm T}}

\newcommand{\paralleltransport}{{\rm P}}

\newcommand{\metric}{{\rm g}}

\newcommand{\bxi}{\boldsymbol{ \xi}}

\newcommand{\Tr}{{\rm Tr}}

\begin{document}

\title{\textbf{Canonical divergence for measuring classical and quantum complexity}}
\author{Domenico Felice}
\email{felice@mis.mpg.de}
\affiliation{Max Planck Institute for Mathematics in the Sciences\\
 Inselstrasse 22--04103 Leipzig,
 Germany}
 \author{Stefano Mancini}
 \email{stefano.mancini@unicam.it}
 \affiliation{School of Science and Technology University of Camerino, Camerino, Italy\\
 INFN-Sezione di Perugia, Perugia, Italy}
 \author{Nihat Ay}
\email{nay@mis.mpg.de}
\affiliation{ Max Planck Institute for Mathematics in the Sciences\\
 Inselstrasse 22--04103 Leipzig,
 Germany\\
Santa Fe Institute, 1399 Hyde Park Rd, Santa Fe, NM 87501, USA\\
 Faculty of Mathematics and Computer Science, University of Leipzig, PF 100920, 04009 Leipzig, Germany}

\begin{abstract}
A new canonical divergence is put forward for generalizing an information-geometric measure of complexity for both, classical and quantum systems. On the simplex of probability measures it is proved that the new divergence coincides with the Kullback-Leibler divergence, which is used to quantify how much a probability measure deviates from the non-interacting states that are modeled by exponential families of probabilities. On the space of positive density operators, we prove that the same divergence reduces to the quantum relative entropy, which quantifies many-party correlations of a quantum state from a Gibbs family.
\end{abstract}


\maketitle

\section{Introduction}

{The many fields} 
of applicability of methods of
information geometry to the science of complexity encompass both classical {and quantum systems} \cite{FeliceChaos}. 
Among them, an information-geometric approach to the complexity as the extent to which an object, as a whole, is more than its parts was established in \cite{Aycomplexity} and then {developed to relate various known measures of complexity to a general class of information-geometric complexity measures} (see \cite{Ay17} for a comprehensive {overview} on this topic). The general idea for quantifying the extent to which the system is more than the sum of its parts is the following. Let $\mathcal{S}$ be a set of systems; for any system $S\in\mathcal{S}$, we assign the collection of system parts which may be an element of a set $\mathcal{S}_0$ that formally differs from $\mathcal{S}$. The corresponding assignment $\Pi:\mathcal{S}\rightarrow\mathcal{S}_0$ can be interpreted as a reduced description of the system $S$ in terms of its parts. Having the parts $\Pi(S)$, we have to reconstruct $S$ by taking the sum of the parts in order to obtain a system that can be compared with the original system. The corresponding construction map is denoted by $\Sigma:\mathcal{S}_0\rightarrow\mathcal{S}$. The composition
$$
\mathrm{P}(S):=(\Sigma\circ\Pi)(S)
$$
then corresponds to the sum of parts of the system $S$, and we can compare $S$ with $\mathrm{P}(S)$. It turns out that $\mathrm{P}${, under natural conditions,} is the projection $\mathrm{P}:\mathcal{S}\rightarrow\mathcal{N}$ to the set of non-complex systems $\mathcal{N}:=\{S\in\mathcal{S}\,|\,\mathrm{P}(S)=S\}$ \cite{Ay11}. Therefore, the quantification of how much the system $S$ differs from $\mathrm{P}(S)$ is established by a divergence function $\Div:\mathcal{S}\times\mathcal{S}\rightarrow\RR$ such that
\begin{equation}
\label{divergence}
\Div(S,S^{\prime})\geq 0,\qquad \Div(S,S^{\prime})=0\quad \mbox{iff}\quad S=S^{\prime}\,.
\end{equation}

Finally, the complexity of a system $S$ is defined by
\begin{equation}
\label{complexity}
\mathrm{C}(S):=\Div(S,\mathrm{P}(S))\,.
\end{equation}

{Clearly, there are many choices for the divergence $\Div$, thus such a complexity measure is far from being unique. However,   to ensure compatibility with $\mathrm{P}$, one has to further assume that $\Div$ satisfies
\begin{equation}
\label{compatibility}
\mathrm{C}(S)=\Div(S,\mathrm{P}(S))=\inf_{S^{\prime}\in\mathcal{N}}\Div(S,S^{\prime})\,.
\end{equation}

Here comes the role of a canonical divergence for providing an information-geometric measure of complexity which can be interpreted as unique. 

In the {framework} of information geometry, a dual structure $(\metric,\nabla,\nabla^*)$ on a smooth manifold $\Ma$ is given in terms of a metric tensor and two affine connections, which are dual in the following sense \cite{Amari00}:
$$
X\,\metric\left(Y,Z\right)=\metric\left(\nabla_X Y,Z\right)+\metric\left(Y,\nabla^*_X Z\right),\quad \forall\, X,Y,Z\,\in\Tau(\Ma)\,,
$$
where $\Tau(\Ma)$ denotes the space of sections on $\Ma$. Eguchi named a function $\Div:\Ma\times\Ma\rightarrow\RR$ satisfying the property in Equation               \eqref{divergence} as a contrast (or divergence) function whenever $\Div$ allows   recovering  the dual structure $(\metric,\nabla,\nabla^*)$ on $\Ma$ in the following way \cite{Eguchi85}:
\begin{align}\label{gfromDiv}
& \metric_{ij}(p)=-\left.\partial_i\partial_j^{\prime} \Div(\bxi_p,\bxi_q)\right|_{p=q}=\left.\partial^{\prime}_i\partial_j^{\prime} \Div(\bxi_p,\bxi_q)\right|_{p=q}\\
\label{GfromDiv}
& \Gamma_{ijk}(p)=-\left.\partial_i\partial_j\partial_k^{\prime} \Div(\bxi_p,\bxi_q)\right|_{p=q}, \qquad {\Gamma}^*_{ijk}(p)=-\left.\partial^{\prime}_i\partial^{\prime}_j\partial_k \Div(\bxi_p,\bxi_q)\right|_{p=q}\ ,
\end{align}
where 
\begin{equation*}
\partial_i=\frac{\partial}{\partial \xi_p^i} \quad \mbox{and}\quad \partial^{\prime}_i=\frac{\partial}{\partial \xi_q^i}
\end{equation*} 
and $\{\bxi_p:=(\xi_p^1,\ldots,\xi_p^n)\}$ and $\{\bxi_q:=(\xi_q^1,\ldots,\xi_q^n)\}$ are local coordinate systems of $p$ and $q$, respectively. 
Here, $\Gamma_{ijk}=\metric\left(\nabla_{\partial_i}\partial_j,\partial_k\right)$ and $\Gamma^*_{ijk}=\metric\left(\nabla^*_{\partial_i}\partial_j,\partial_k\right)$ are the connection symbols of $\nabla$ and $\nabla^*$, respectively. 
{The investigation on a divergence function allowing to recover the dualistic structure on a smooth manifold is usually referred to as the inverse problem in information geometry. 
{Matumoto} \cite{matumoto1993} showed that such divergence exists for any statistical manifold.} However, it is not unique and there are infinitely many divergences that give the same dual structure. Hence,     the search for a divergence  that  can be somehow considered as {\it the most natural} {is of upmost importance.} When a manifold is dually flat,  Amari and Nagaoka \cite{Amari00} introduced a Bregman type divergence to this end, with relevant properties concerning the generalized Pythagorean theorem and the geodesic projection theorem.
This is referred to as {\it canonical divergence} and it is commonly assessed as the natural solution of the inverse problem in information geometry for dually flat manifolds. {However, the need for a general canonical divergence, which applies to any dualistic structure, is a very crucial issue, as pointed out in \cite{AyTusch}. In any case, such a divergence should recover the canonical divergence of Bregman type if applied to a dually flat structure.} In addition, in the self-dual case where $\nabla=\nabla^*$ coincides with the Levi--Civita connection of $\metric$, the divergence $\Div$ should be one half of the squared Riemannian distance: $\Div(p,q)=\frac{1}{2}\,d(p,q)^2$ \cite{Ay17}. In the context of the information-geometric approach to complexity, a further requirement is {{needed}}   to ensure the compatibility in Equation               \eqref{compatibility}. This is the geodesic projection property, which, in the present context, states that every minimizer $\mathrm{P}(S)$ of $\Div$ is achieved by the geodesic projection of $S$ onto the set of non-complex systems. In \cite{Ay15}, Ay and Amari   recently introduced a canonical divergence that satisfies all these requirements. Such a divergence is defined in terms of geodesic integration of the inverse exponential map. More precisely, given $p,q\in\Ma$ and the $\nabla$-geodesic $\widetilde{\sigma}(t)\,(0\leq t\leq 1)$ connecting $q$ with $p$,  the canonical divergence introduced in \cite{Ay15} {is given by}
\begin{equation}
\label{AyDiv}
\Div(p,q):=\int_0^1\left\langle\nihat_t(p),\dot{\widetilde{\sigma}}(t)\right\rangle_{\widetilde{\sigma}(t)}\,dt\,,\quad \nihat_t(p):=\exp_{\widetilde{\sigma}(t)}^{-1}(p)\,.
\end{equation}

Here, $\exp:\tangent\Ma\rightarrow\Ma$ denotes the exponential map of $\nabla$, which is defined by $\exp(X)=\sigma_X(1)$ whenever the $\nabla$-geodesic $\sigma_X(t)$, {satisfying $\dot{\sigma}_X(0)=X$,} exists on an interval of $t$ containing $[0,1]$. Therefore, if $\sigma(t)\,(0\leq t\leq 1)$ is the $\nabla$-geodesic such that $\sigma(0)=p$ and $\sigma(1)=q$, then $\exp_p^{-1}(q):=\dot{\sigma}(0)$. According to this definition, we have that $\nihat_t(p)=\paralleltransport_{\sigma(t)}\,\nihat_p(\sigma(t))=t\,\dot{\sigma}(t)$, where $\paralleltransport$ is the $\nabla$-parallel transport from $p$ to $\sigma(t)$. This implies that the divergence $\Div(p,q)$ assumes the following useful expression:
\begin{equation}
\label{AyDivnice}
\Div(p,q)=\int_0^1\,t\,\|\dot{\sigma}(t)\|^2\,d t\,.
\end{equation}

Analogously, the dual function of $\Div(p,q)$ is defined as the $\nabla^*$-geodesic integration of the inverse of the $\nabla^*$-exponential map \cite{Ay15}. Therefore, we have for the dual divergence $\Div^*$ a similar expression as Equation               \eqref{AyDivnice} for the canonical divergence $\Div$:
\begin{equation}
\label{AyDivnice*}
\Div^*(p,q)=\int_0^1\,t\,\|\dot{\sigma}^*(t)\|^2\,d t\,,
\end{equation}
where $\sigma^*(t)\,(0\leq t\leq 1)$ is the $\nabla^*$-geodesic connecting $p$ with $q$. Therefore, the compatibility in Equation               \eqref{compatibility} of $\Div$ with $\mathrm{P}$ suggests that the projection $\mathrm{P}(S)$ of a system $S$ onto the space of non-complex systems can be achieved along the geodesic connection $S$ with $\mathrm{P}(S)$. {Actually, it has recently been proved that the $\nabla$-geodesic minimizes the action integral of a suitably chosen kinetic energy \cite{HJ2}. An analogous result {{holds}} about the $\nabla^*$-geodesic. In this way, both divergences, $\Div(p,q)$ and $\Div^*(p,q)$, turn out to solve the Hamilton--Jacobi problem in information geometry, as put forward in \cite{HJ1}.}

The search for a general canonical divergence is still an open problem and it turns out to be of upmost importance in the context of the information-geometric approach to complexity (see progresses along this avenue put forward in \cite{Ay15,Felice18}).

{In this article, }
we aim to propose the canonical divergence in      Equation               \eqref{AyDivnice} as an efficient tool for providing a unified definition of complexity measures. For this reason, we firstly consider $\Div$ on the simplex of probability distributions where a measure of complexity as one instance of Equation               \eqref{complexity} is supplied in terms of the Kullback--Leibler (KL)-divergence \cite{Ay11}.

The general methods described for defining the complexity measure in Equation               \eqref{complexity} can be particularized to the systems consisting of a finite node set $V$ and each node $v\in V$ can be in finitely many states $I_v$. Then,     we model the whole system as a probability measure $p$ on the corresponding product configuration set $I_V=\prod_{v\in V} I_v$. The parts are given by marginals $p_A$ where $A$ is taken from a {set of subsets of $V$, denoted by $\mathfrak{S}$}. Therefore, the decomposition map $\Pi$ reads in this case as $\Pi(p)=\left(p_A\right)_{A\in\mathfrak{S}}$, whereas  the reconstruction map $\Sigma$ is defined by the maximum entropy estimate $\hat{p}$ of $p$, {leading to the} projection $\pi_{\mathfrak{S}}:p\mapsto\hat{p}$. {The image of $\pi_{\mathfrak{S}}$ turns out to be the closure of an exponential family $\mathcal{E}_{\mathfrak{S}}$, which plays the role of the set $\mathcal{N}$ of non-complex systems}. A deviation measure, which is compatible with the maximum entropy projection $\pi_{\mathfrak{S}}$ is then the {(KL)}-divergence, which is defined by
\begin{equation}
\label{K-Ldivergence}
\KL(p,q):=\sum_{i=1}^{n+1}\,p_i\,\log\left(\frac{p_i}{q_i}\right)
\end{equation}
on the $n$-simplex $\mathcal{P}_n=\{p=(p_1,\ldots,p_{n})\,|\,p_i> 0\,,\sum_i p_i=1\}$ \cite{Eguchi85}. Finally, the measure of complexity as one instance of Equation               \eqref{complexity} is obtained by
\begin{equation}
\label{KLcomplexity}
\KL\left(p,\mathcal{E}_{\mathfrak{S}}\right):=\inf_{q\in\mathcal{E}_{\mathfrak{S}}}\KL(p,q)=\KL(p,\hat{p})\,.
\end{equation}

{ We may notice that, if $\mathfrak{S}$ consists of all subsets of $V$ of cardinality $1$, elements of the set $\mathcal{E}_{\mathfrak{S}}$ of non-complex systems  are totally uncorrelated in the sense that $q\in \mathcal{E}_{\mathfrak{S}}$ has the product form $q=q_1\otimes\ldots\otimes q_n$ \cite{Aycomplexity}. Consider random variables $X_1,\ldots,X_n$   with joint probability distribution $p$ and marginal probability distributions $p_1,\ldots,p_n$. Then, we have 
$$
\KL\left(p,\mathcal{E}_{\mathfrak{S}}\right)=\KL\left(p,p_1\otimes\ldots\otimes p_n\right)=\sum_i H(X_i)-H(X_1,\ldots,X_n)\,,
$$
where $H$ is the Shannon entropy. This quantity is referred to as {\it multi information} and denoted by $I(X_1,\ldots,X_n)$. In particular, when $n=2$, this is nothing but the {\it mutual information}.}
Very remarkably, {the minimizer $\hat{p}$ in the closure of $\mathcal{E}_{\mathfrak{S}}$ of the (KL)-divergence, namely $\KL(p,\hat{p})=\inf_{q\in\mathcal{E}_{\mathfrak{S}}}\KL(p,q)$, is obtained by projecting $p$ onto the closure of $\mathcal{E}_{\mathfrak{S}}$ along a mixture $(m)$-geodesic} \cite{Eguchi92}. This is usually referred to as the geodesic projection property of the (KL)-divergence. The geometric structure given by the Fisher metric, the mixture $(m)$ and exponential $(e)$ affine connections was introduced by Amari and Nagaoka on the space of probability densities for studying statistical estimation problems \cite{Amari00}. 

{In this article, }
we then consider both divergences, $\Div$ and $\Div^*$, on $\mathcal{P}_n$ with the endowed dualistic structure given by the classic Fisher metric and the mixture $(m)$ and the exponential $(e)$ connections. Here, we show that $\Div(q,p)=\KL(q,p)=\Div^*(p,q)$. Actually, this result has already { been}   shown in \cite{Ay15}. However, we prove it differently by relying on the nice representations of $\Div$ and $\Div^*$ given by Equations               \eqref{AyDivnice} and                 \eqref{AyDivnice*}, respectively. This proves that $\Div$ can be interpreted as a generalization of the (KL)-divergence.

 A further step for proving the effectiveness of $\Div$ is to consider it (and its dual function) on the manifold of finite quantum states where the general idea for defining a complexity measure of a classic system expressed by Equation               \eqref{complexity} has been extended to the quantum setting in terms of the quantum relative entropy \cite{Weis15}. More precisely, { by considering a composite set of $n\in\NN$ {units (parties, particles)}, 
 $[n]:=\{1,\ldots,n\}$, the composite system is described by the product algebra $\mathcal{A}_{[n]}:=\mathcal{A}_1\otimes\ldots\otimes\mathcal{A}_n$. Here, $\mathcal{A}_i\subset M_{n_i}$ is the $C^*$-subalgebra of complex $n_i\times n_i$ matrices such that the identity $\mathbb{I}_{n_i}\in\mathcal{A}_i$. The many-party correlations are quantified in the state of a composite quantum system which can not be observed in subsystems composed of less than a given number of parties. In this context, the exponential families, which amount to the non-complex system in the classical case, are replaced by states  that  are fully described by their restriction to selected subsystems. These correspond to the family of Gibbs states $\mathcal{E}_k:=\{e^{H_k}/\Tr e^{H_k}\}$ of the $k$-local Hamiltonians $H_k$. Here, a $k$-local Hamiltonian is defined as a sum of product terms $a_1\otimes\ldots\otimes a_n$ with at most $k$-non-scalar factors $a_i$, where $a_i$ denotes a real self-adjoint operator. Therefore, the many-party correlations of a composite quantum state $\rho\in\mathcal{A}_{[n]}$ which captures all correlations in $\rho$ that cannot be observed in any $k$-party subsystem is the divergence
\begin{equation}
\label{manyparty}
\Q(\rho,\mathcal{E}_k):=\inf_{\sigma\in\mathcal{E}_k}\Q(\rho,\sigma)
\end{equation}
from the Gibbs family $\mathcal{E}_k$ \cite{Weis15}. Here, the divergence $\Q(\rho,\sigma)$ is the quantum relative entropy defined by
\begin{equation}
\label{quantumrelativentropy}
\Q(\rho,\sigma)=\Tr\,\rho\left(\log\rho-\log\sigma\right)\,,
\end{equation}
where $\Tr$ denotes the trace operator on the finite-dimensional Hilbert space of density matrices.  Similar   to the classical case, we can consider the family $\mathcal{E}_1$ of Gibbs states whose closure corresponds to the set of product states $\sigma_1\otimes\ldots\otimes\sigma_n$. Consider then a composite quantum state $\rho\in\mathcal{A}_{[n]}$ such that
$$
\Tr\left(\sigma_i\,a\right)=\Tr\left(\rho\,(a\otimes \mathbb{I}_{[n]\backslash \{i\}})\right)\,,
$$ 
where $a\in\mathcal{A}_{\{i\}}=\mathcal{A}_i$ and $\mathbb{I}_{[n]\backslash \{i\}}$ is the identity operator on the product $\mathcal{A}_1\otimes\ldots\hat{\mathcal{A}}_i\ldots\otimes \mathcal{A}_n$  where $\mathcal{A}_i$ is missing. In this case, the many-party correlations of $\rho$ is the {\it quantum multi information}:
$$
\Q(\rho,\mathcal{E}_1)=\sum_i\widetilde{H}(\sigma_i)-\widetilde{H}(\rho)\,,
$$
where $\widetilde{H}(\rho)=-\Tr(\rho\log\rho)$ is the von Neumann entropy of $\rho$.
In particular, when $n=2$, this corresponds to the {\it quantum mutual information}.} Algorithms for the evaluation of $\Q(\rho,\mathcal{E}_k)$ as a complexity measure for quantum states  are  studied in \cite{Niekamp13}. In that context, the many-party correlations is related to the entanglement of quantum systems as defined in \cite{Vedral97}.

The scope of the present article is mainly to present the canonical divergence $\Div$ defined in Equation               \eqref{AyDivnice} as an important tool for generalizing the concept of complexity measure claimed by Equation               \eqref{KLcomplexity} for classical systems as well as the concept of many-party correlation given by Equation               \eqref{manyparty} for quantum systems. To this end, we consider the space of density matrices endowed with the quantum analog of the Fisher metric and the mixture $(m)$ and exponential $(e)$ affine connections. {This structure turns out to be induced on the manifold of positive density operators by the Bogoliubov inner product \cite{Nagaoka95}}. In this setting, we prove that the divergence introduced in \cite{Ay15} reduces to the quantum relative entropy.  In addition, we also show that $\Div(\sigma,\rho)=\Q(\sigma,\rho)=\Div^*(\rho,\sigma)$.

The layout of the paper is as follows.   Section \ref{Section2} is devoted to the calculation of the canonical divergence and its dual function on the simplex of probability distributions. In Section \ref{Section3}, we describe the differential geometrical framework for finite quantum systems induced by the Bogoliubov inner product. In this particular framework, we then prove that the divergence given by Equation               \eqref{AyDivnice} reduces to the quantum relative entropy.  Finally, we draw some conclusions in Section \ref{Section4} by outlining the results obtained in this work and discussing possible extensions.
 
\section{Canonical Divergence on the Simplex of Probability Measures}\label{Section2}

 A dualistic structure on the simplex of probability measures was introduced by Amari in terms of the Fisher metric, the mixture $(m)$ and exponential $(e)$ connections \cite{Amari}. Given a finite set $I=\{1,\ldots,n\}$, we can represent probability measures on the set $I$ as elements of $\RR^n$. In this representation, the Dirac measures $\delta^i,\,i=1,\ldots,n$ form the canonical basis of $\RR^n$. Then,     the $(n-1)$-dimensional simplex of probability measure is given by
\begin{equation}
\label{simplex}
\esse_n:=\left\{p=\sum_i\,p_i\delta^i\in\RR^n\,|\,p_i>0\,\mbox{for all}	\,i,\,\mbox{and}\,\sum_i\,p_i=1\right\}\,.
\end{equation}

In this section, we show that the canonical divergence $\Div(p,q)$ coincides with the Kullback--Leibler divergence whenever $p,q\in\esse_n$. In addition, we prove that, for the dual canonical divergence, the following relation $\Div^*(p,q)=\KL(q,p)$ holds true. According to Equations               \eqref{AyDivnice} and                 \eqref{AyDivnice*}, we need the Fisher metric defined on the tangent bundle $\tangent\esse_n$, the mixture $(m)$-geodesic and the exponential $(e)$-geodesic both connecting $p$ with $q$. On the tangent space $\tangent_p\esse_n$, the Fisher metric results in
\begin{equation}
\label{Fisherinnerproduct}
\metric_p(X,Y):=\sum_i\,\frac{1}{p_i}\,X^i\,Y^i,\qquad X,Y\in\tangent_p\esse_n\,.
\end{equation}

The dualistic structure $(\metric,\nabla,\nabla^*)$ on $\esse_n$, given by the Fisher metric, the $(m)$-connection $\nabla$ and the $(e)$-connection $\nabla^*$, is dually flat, and the $(m)$- and $(e)$-geodesics connecting $p$ with $q$ are \cite{Ay17}:
\begin{align}
\label{mclassic}
& \gamma_m(t)=p+t(q-p),\quad t\in[0,1]\\
\label{eclassic}
& \gamma_e(t)=\sum_i\frac{p_i\left(\frac{q_i}{p_i}\right)^t}{\sum_jp_j\left(\frac{q_j}{p_j}\right)^t}\,\delta^i,\quad t\in[0,1]\,.
\end{align}

\vspace{.2cm}

  We are now ready to compute the canonical divergence $\Div(p,q)$ for arbitrary $p,q\in\esse_n$. From Equations               \eqref{AyDivnice},               \eqref{Fisherinnerproduct},  and               \eqref{mclassic}, we have that
\begin{eqnarray}
\Div(p,q)&=& \int_0^1\,t\|\dot{\gamma}_m(t)\|_{\gamma_m(t)}^2\,dt\nonumber\\\nonumber
&=& \sum_i\int_0^1\,t\frac{1}{p_i+t(q_i-p_i)}(q_i-p_i)^2\,dt\\\nonumber
&=& \sum_i\left(q_i-p_i+p_i\log\frac{p_i}{q_i}\right)\nonumber\\
&=& \sum_i\,p_i\log\frac{p_i}{q_i}=\KL(p,q)\,,
\end{eqnarray}
where we use  $\sum_i(q_i-p_i)=0$ because $p,q\in\esse_n$. Analogously, we can compute the dual canonical divergence $\Div^*(p,q)$ by means of Equation               \eqref{AyDivnice*}. Therefore, by using Equations               \eqref{Fisherinnerproduct} and                 \eqref{eclassic}, we obtain that
\begin{eqnarray}
\Div^*(p,q) &=& \int_0^1\,t\|\dot{\gamma}_e(t)\|_{\gamma_e(t)}^2\,dt\nonumber\\\label{edivcomp}
&=& \sum_i\int_0^1\,t\dot{\gamma}_e^i(t)\,\frac{\dot{\gamma}_e^i(t)}{\gamma_e^i(t)}\,dt\,.
\end{eqnarray}

To develop further the calculation, let us {analyze} the derivative $\dot{\gamma}_e^i(t)$. Recall that 
$$
\gamma_e^i(t)=\frac{p_i\left(\frac{q_i}{p_i}\right)^t}{\sum_jp_j\left(\frac{q_j}{p_j}\right)^t}\,.
$$

Therefore, by taking the derivative of $\gamma_e^i(t)$ with respect to $t$, we obtain
\begin{eqnarray*}
\dot{\gamma}_e^i(t) &=& \frac{p_i\left(\frac{q_i}{p_i}\right)^t\log\frac{q_i}{p_i}}{\sum_jp_j\left(\frac{q_j}{p_j}\right)^t} - p_i\left(\frac{q_i}{p_i}\right)^t\frac{\sum_jp_j\left(\frac{q_j}{p_j}\right)^t\log\frac{q_j}{p_j}}{\left(\sum_jp_j\left(\frac{q_j}{p_j}\right)^t\right)^2}\\
\\
&=& \gamma_e^i(t)\left(\log\frac{q_i}{p_i}-\frac{\sum_jp_j\left(\frac{q_j}{p_j}\right)^t\log\frac{q_j}{p_j}}{\sum_jp_j\left(\frac{q_j}{p_j}\right)^t}\right)\\
\\
&=&\gamma_e^i(t)\left(\log\frac{q_i}{p_i}-\frac{\mathrm{d}}{\mathrm{d}t}\log\sum_jp_j\left(\frac{q_j}{p_j}\right)^t\right)\,.
\end{eqnarray*}

By stepping back to Equation               \eqref{edivcomp}, we start by performing an integration by parts:
\begin{eqnarray}
\Div^*(p,q) &=& \sum_i\left(\left[\gamma_e^i(t)\left(t\frac{\dot{\gamma}_e^i(t)}{\gamma_e^i(t)}\right)\right]_0^1-\int_0^1\gamma_e^i(t)\frac{\dot{\gamma}_e^i(t)}{\gamma_e^i(t)}\,dt+\int_0^1\,t\frac{\mathrm{d^2}}{\mathrm{d}t^2}\log\sum_jp_j\left(\frac{q_j}{p_j}\right)^t\,dt\right)\,,\label{divbyparts}
\end{eqnarray}
where the last term is obtained by noticing that
$$
\frac{\dot{\gamma}_e^i(t)}{\gamma_e^i(t)}=\left(\log\frac{q_i}{p_i}-\frac{\mathrm{d}}{\mathrm{d}t}\log\sum_jp_j\left(\frac{q_j}{p_j}\right)^t\right)\,.
$$

Since we know that 
$$\frac{\mathrm{d}}{\mathrm{d}t}\log\sum_jp_j\left(\frac{q_j}{p_j}\right)^t=\frac{\sum_jp_j\left(\frac{q_j}{p_j}\right)^t\log\frac{q_j}{p_j}}{\sum_jp_j\left(\frac{q_j}{p_j}\right)^t}\,,
$$
we can observe that $\frac{\dot{\gamma}_e^i(1)}{\gamma_e^i(1)}=\left(\log\frac{q_i}{p_i}-\sum_jq_j\log\frac{q_j}{p_j}\right)$. Hence, we obtain from Equation               \eqref{divbyparts}
\begin{eqnarray*}
\Div^*(p,q)&=&\sum_i\Bigg(q_i\left(\log\frac{q_i}{p_i}-\sum_jq_j\log\frac{q_j}{p_j}\right)-\left[\gamma_e^i(t)\right]_0^1+\left[t\frac{\mathrm{d}}{\mathrm{d}t}\log\sum_jp_j\left(\frac{q_j}{p_j}\right)^t\right]^1_0\\
&&-\int_0^1\frac{\mathrm{d}}{\mathrm{d}t}\log\sum_jp_j\left(\frac{q_j}{p_j}\right)^t\,dt\Bigg)\\
&=& \sum_i\left(q_i\log\frac{q_i}{p_i}-q_i\sum_jq_j\log\frac{q_j}{p_j}-q_i+p_i+\sum_jq_j\log\frac{q_j}{p_j}-\left[\log\sum_jp_j\left(\frac{q_j}{p_j}\right)^t\right]_0^1\right)\\
&=& \sum_i q_i\log\frac{q_i}{p_i}-\sum_iq_i\sum_jq_j\log\frac{q_j}{p_j}-\sum_iq_i+\sum_ip_i+\sum_jq_j\log\frac{q_j}{p_j}-\log\sum_jq_j+\log\sum_jp_j\\
&=& \sum_i q_i\log\frac{q_i}{p_i}\,,
\end{eqnarray*}
because $p,q\in\esse_n$.
This proves that
$$
\Div^*(p,q)=\KL(q,p)=\Div(q,p)\,.
$$






\section{Geometric Structure of a Manifold of Quantum States}\label{Section3}
  
  We start this section by showing that natural analogs of the Fisher metric and the exponential and mixture connections are defined on a manifold of quantum states \cite{Nagaoka95}. To this end, we need to specify an inner product on the space of density operators. Since the divergence $\Div$ of Equation               \eqref{AyDivnice} is defined on a statistical manifold $(\Ma,\metric,\nabla,\nabla^*)$ with symmetric connections, we choose the Bogoliubov inner product. This is because of a well-known result that claims   the $(e)$-connection  induced by a generalized covariance is symmetric if and only if such a covariance is the Bogoliubov inner product \cite{Amari00}. {At the end of this section, we   motivate this choice in more detail.}

  Let $\Hilb$ be a finite-dimensional Hilbert space, $\mathcal{A}=\{A\,|\,A=A^*\}$ be the space of all the Hermitian operators on $\Hilb$ and $\mathcal{S}=\{\rho\,|\,\rho=\rho^*>0,\,\Tr\rho=1\}$ be the space of positive density operators on $\Hilb$. Since $\mathcal{S}$ is an open subset of $\mathcal{A}_1:=\{A\,|\,A=A^*,\,\Tr A=1\}$, then it can be naturally seen as a smooth manifold of dimension $n=\left(\dim\Hilb\right)^2-1$ \cite{Nagaoka95}. Let $D\in\tangent_{\rho}\mathcal{S}$ be a tangent vector at $\rho$ to $\mathcal{S}$; we call $D^{(m)}\in\mathcal{A}_0:=\{A\,|\,A\in\mathcal{A},\,\Tr A=0\}$ its $(m)$-representation and symbolically write 
\begin{equation}
\label{m-representation}
D^{(m)}=D\rho\,.
\end{equation}

{It is worth noticing that, as an element of the tangent space, $D$  can be naturally interpreted as a derivative. As an example, when a coordinate system $\{\theta^i\}$ is given on $\mathcal{S}$ so that each state is parameterized as $\rho\equiv\rho_{\theta}$, the $(m)$-representation of the natural basis vector is written as $(\partial_i)^{(m)}=\partial_i\rho_{\theta}$, where $D=\partial_i=\partial/\partial \theta^i$.}
This allows us to introduce the $(m)$-connection on the manifold $\mathcal{S}$ of the quantum states in terms of the covariant derivative $\nabla^{(m)}:\Tau(\mathcal{S})\times\Tau(\mathcal{S})\rightarrow\Tau(\mathcal{S})$, which is defined by the following relation:
\begin{equation}
\label{m-connection}
\left(\nabla^{(m)}_X Y\right)^{(m)}=X\left(Y^{(m)}\right),\quad \forall\, X,\,Y\,\in\Tau(\esse)\,,
\end{equation}
where the right hand side means the derivative by $X$ of $Y^{(m)}:\esse\rightarrow\mathcal{A}_0$ and $\Tau(\esse)$ denotes the space of sections on $\esse$.

     To introduce the $(e)$-connection on $\esse$, we need to specify a family $\{\langle\cdot,\cdot\rangle_{\rho}\,|\,\rho\in\esse\}$ of inner products on $\mathcal{A}$ usually named as {\it generalized covariance}. For the reason   mentioned above, we consider the Bogoliubov inner product, which is given by
\begin{equation}
\label{Bogoliubov}
\left\langle A,B\right\rangle_{\rho}:=\int_0^1\,\Tr\left(\rho^{\lambda}A\rho^{1-\lambda}B\right)\,d\lambda\,,\quad\forall\, A,B\in\mathcal{A}\,.
\end{equation}

Given $D\in\tangent_{\rho}\esse$, we then define the $(e)$-representation of $D$ as the Hermitian operator $D^{(e)}\in\mathcal{A}$ satisfying the following relation:
\begin{equation}
\label{e-representation}
\Tr\left(D^{(m)}\,A\right)=:\left\langle D^{(e)},A\right\rangle_{\rho}=\int_0^1\,\Tr\left(\rho^{\lambda}D^{(e)}\rho^{1-\lambda}A\right)\,d\lambda\,,\quad\forall\, A\in\mathcal{A}\,.
\end{equation}

{ For all $A\in\mathcal{A}$, we assume $\langle A,\mathbb{I}\rangle_{\rho}=\langle A\rangle_{\rho}=\Tr(\rho A)$ ($\mathbb{I}$ denotes the identity operator). Thus, we can see that the derivative of the function $\langle A\rangle:\rho\rightarrow\langle A\rangle_{\rho}$ by $D$ is written as
$$
D\langle A\rangle=\Tr(D^{(m)}A)=\langle D^{(e)},A\rangle_{\rho}\,.
$$}

This implies that { we can consider the $(e)$-representation $D^{(e)}\in\mathcal{A}$ of a given $D\in\tangent_{\rho}\esse$ as 
\begin{equation}
\label{evsm-representation}
D\rho=\int_0^1\,\rho^{\lambda}\,D^{(e)}\,\rho^{1-\lambda}\,d\lambda\,.
\end{equation}}

Therefore, it turns out that $D^{(e)}$ is the derivative of the map $\rho\mapsto\log\rho$ from $\esse$ to $\mathcal{A}$, which may be written as follows:
\begin{equation}
\label{Boge-representation}
D^{(e)}=D\log\rho\,.
\end{equation}

By {considering} $\left\langle D^{(e)},\mathbb{I}\right\rangle_{\rho}=\left\langle D^{(e)}\right\rangle_{\rho}=\Tr\left(\rho\, D^{(e)}\right)$,  we can immediately observe that
$$
\left\langle D^{(e)}\right\rangle_{\rho}=D\left\langle\mathbb{I}\right\rangle_{\rho}=0\,.
$$

This proves that, although the $(e)$-representation depends on the choice of the generalized covariance, the space $\tangent_{\rho}^{(e)}\esse:=\{D^{(e)}\,|\, D\in\tangent_{\rho}\esse\}$ can be simply written as follows
\begin{equation}
\label{e-tangentspace}
\tangent_{\rho}^{(e)}\esse=\{A\,|\, A\in\mathcal{A},\,\left\langle A\right\rangle_{\rho}=\Tr\left(\rho\,A\right)=0\}\,.
\end{equation} 

This fact supplies the manifold $\esse$ of quantum states with the $(e)$-connection. To see this, let us consider the linear isomorphism $D\mapsto D^{\prime}$ from $\tangent_{\rho}\esse$ to $\tangent_{\rho^{\prime}}\esse$  defined by $D^{\prime(e)}=D^{(e)}-\left\langle D^{(e)}\right\rangle_{\rho^{\prime}}$. By writing this correspondence as $D^{\prime}=\left[D\right]_{\rho^{\prime}}$, $D=\left[D^{\prime}\right]_{\rho}$, the $(e)$-connection $\nabla^{(e)}$ is then defined by
\begin{equation}
\label{e-connection}
\left(\nabla_X^{(e)}Y\right)_{\rho}=X_{\rho}\left[Y\right]_{\rho},\quad \forall\,\rho\in\esse,\,\forall X,Y\in\Tau(\esse)\,,
\end{equation}
where the right hand side means the derivative by $X_{\rho}$ of $\left[Y\right]_{\rho}:\esse\rightarrow\tangent_{\rho}\esse$. 

Finally, we define the inner product $\metric_{\rho}$ on $\tangent_{\rho}\esse$ by
\begin{equation}
\label{QuantumFisher}
\metric_{\rho}\left(X,Y\right):=\left\langle X^{(e)},Y^{(e)}\right\rangle_{\rho}=\Tr\left(X^{(m)}\,Y^{(e)}\right)\,,
\end{equation}
which is usually called the {\it quantum Fisher metric}. The procedure thus far described endows the manifold $\esse$ of quantum states with a geometric structure $(\metric,\nabla^{(e)},\nabla^{(m)})$ given by the quantum Fisher metric, and two torsion-free connections, namely the $(e)$-connection $\nabla^{(e)}$ and the $(m)$-connection $\nabla^{(m)}$, which are dual with respect to $\metric$ in the following sense:
\begin{equation}
\label{dualFisher}
X\,\metric\left(Y,Z\right)=\metric\left(\nabla_X^{(m)}Y,Z\right)+\metric\left(Y,\nabla^{(e)}_XZ\right)\,,\quad \forall\, X,Y,Z\in\Tau(\esse)\,.
\end{equation}

In addition, the dual structure $(\metric,\nabla^{(m)},\nabla^{(e)})$ is dually flat, meaning that the curvature tensors of $\nabla^{(e)}$ and $\nabla^{(m)}$ are both null.

Suppose that a coordinate system $\{\xi_i\}$ is given and that each element $\rho\in\esse$ is specified by the coordinate $\bxi\in\RR^n$ as $\rho\equiv \rho_{\bxi}$. According to Equation               \eqref{m-representation}, we have that the mixture representation $\partial_i^{(m)}$ of $\partial_i=\partial/\partial\xi^i$ is given by $\partial^{(m)}_i\rho=\partial_i\rho_{\bxi}$, whereas, by Equation               \eqref{e-representation}, we have that the exponential representation $\partial_i^{(e)}$ of $\partial_i$ is written as $\partial_i^{(e)}\rho=\partial_i\log\rho_{\bxi}$. Therefore, the dual structure $(\metric,\nabla^{(e)},\nabla^{(m)})$ with respect to an arbitrary coordinate system $\{\xi^i\}$ reads as follows
\begin{align}
&g_{ij}= \Tr\left(\partial_i\rho_{\bxi}\,\partial_j\log\rho_{\bxi}\right)\\
&\Gamma^{(e)}_{ijk}=\Tr\left(\partial_i\partial_j\log\rho_{\bxi}\,\partial_k\rho_{\bxi}\right),\quad \Gamma^{(m)}_{ijk}=\Tr\left(\partial_i\partial_j\rho_{\bxi}\,\partial_k\log\rho_{\bxi}\right)\,.
\end{align}

\vspace{.2cm}

  {A generalized covariance is a family $\{\langle\cdot,\cdot\rangle_{\rho}\,|\,\rho\in\esse\}$ of inner products on the space of Hermitian { operators}  $\mathcal{A}$ on the Hilbert space $\Hilb$, where $\langle A,B\rangle_{\rho}$ depends smoothly on $\rho$ for all $A,B\in\mathcal{A}$ and that satisfies the following properties:
\begin{itemize}
\item For every $U$ unitary matrix on the Hilbert space $\Hilb$, it is
$$
\langle U A U^*,U B U^*\rangle_{U\rho U^*}=\langle A, B\rangle_{\rho},\quad \forall\, A, B\in\mathcal{A},\,\rho\in\esse\,.
$$
\item If the Lie bracket $[\rho,A]=0$, then
$$
\langle A,B\rangle_{\rho}=\Tr\left(\rho AB\right)\,.
$$
\end{itemize}

This can be viewed as a quantum version of the $L^2$-product
$$
\langle A,B\rangle_{p}=\mathbb{E}_p[A,B]
$$
of random variables $A$ and $B$ with respect to a probability measure $p$. Since $\mathbb{E}_p[A,B]$ is the covariance of $A$ and $B$ when their expectations vanish, we can call the family $\{\langle\cdot,\cdot\rangle_{\rho}\,|\,\rho\in\esse\}$ satisfying the above conditions a {\it generalized covariance}.

According to the theory by Eguchi, a divergence function $\Div:\Ma\times\Ma\rightarrow\RR^*$ induces a dual structure $(\metric,\nabla,\nabla^*)$ on $\Ma$ in the way expressed by Equations               \eqref{gfromDiv} and                 \eqref{GfromDiv}. It turns out that the connections $\nabla$ and $\nabla^*$ obtained in such a way are torsion-free (or symmetric) \cite{Eguchi92}.      To use the canonical divergence in      Equation               \eqref{AyDivnice} in the quantum setting, we are then forced to select the Bogoliubov inner product for providing the quantum analog of the Fisher metric, the $(m)$-connection and $(e)$-connection on the manifold of positive density operators. Indeed, while the $(m)$-connection is always torsion-free, it turns out that the $(e)$-connection induced on $\esse$ from a generalized covariance is symmetric if and only if such a covariance is the Bogoliubov inner product.
}

\subsection*{{Canonical Divergence on the Manifold} of Quantum States}

 In this section we       show that the divergence function of Equation               \eqref{AyDivnice} reduces to the quantum relative entropy whenever the dual structure $(\metric,\nabla^{(m)},\nabla^{(e)})$ on $\esse$ is given by the Fisher metric (Equation               \eqref{QuantumFisher}), the mixture connection (Equation               \eqref{m-connection}) and the exponential connection (Equation               \eqref{e-connection}).

Let $\rho_1,\rho_2\in\esse$ be two density matrices.      To implement the computation of the divergence $\D(\rho_1,\rho_2)$ for quantum states, we consider the $(m)$-geodesic $\gamma_m(t)=(1-t)\,\rho_1+t\,\rho_2$ \cite{Petz08}. Then, the $(m)$ and $(e)$ representations of the tangent vector $\dot{\gamma}_m(t)$ are easily computed by means of Equations               \eqref{m-representation} and                 \eqref{Boge-representation}, respectively:
\begin{equation}
\label{mixture}
\dot{\gamma}_m^{(m)}(t)=\dot{\gamma}_m(t)=\rho_2-\rho_1,\quad \dot{\gamma}_m^{(e)}(t)=\frac{\mathrm{d}}{\mathrm{d} t}\log\gamma_m(t)\,.
\end{equation}

From Equations               \eqref{AyDivnice} and                 \eqref{QuantumFisher}, we have then
\begin{equation}
\label{AyDivQuantum}
\Div(\rho_1,\rho_2)=\int_0^1\,t\,\Tr\left(\dot{\gamma}_m(t)\,\frac{\mathrm{d}}{\mathrm{d} t}\log\gamma_m(t)\right)\,d t\,.
\end{equation}

Let us recall that $\gamma_m(t)$ is a curve in the space of density matrices and the logarithm of a positive matrix is a well-defined matrix. Therefore, the derivative with respect to $t$ of $\log\gamma_m(t)$ is viewed as the matrix of the derivatives of the entries of $\log\gamma_m(t)$ with respect to $t$. Moreover, the same holds for the integration of a matrix: this is the matrix of the integration of the entries. Finally, since the trace is a linear operator it commutes with the integration. Hence, with the abuse of notation where we keep $\gamma_m$ instead of the entry $(\gamma_m)_{ij}$, the computation in Equation               \eqref{AyDivQuantum} is performed as follows by integration by parts:
\begin{eqnarray*}
\int_0^1\,t\,\dot{\gamma}_m(t)\,\frac{\mathrm{d} }{\mathrm{d} t}\log\gamma_m(t)&=&\left[ t\,\dot{\gamma}_m(t)\,\log\gamma_m(t)\right]_0^1-\int_0^1\,\dot{\gamma}_m(t)\log\gamma_m(t)\,d t\\
&=& (\rho_2-\rho_1)\log\rho_2-\int_{\rho_1}^{\rho_2}\,\log\gamma_m(t)\,d\gamma_m(t)\\
&=&(\rho_2-\rho_1)\log\rho_2-\left[\gamma_m\,\log\gamma_m\right]_{\rho_1}^{\rho_2}\\
&=& \rho_1(\log\rho_1-\log\rho_2)\,.
\end{eqnarray*}

This proves that $\Div(\rho_1,\rho_2)=\Tr\left(\rho_1(\log\rho_1-\log\rho_2)\right)$, which is the quantum relative entropy given by Equation               \eqref{quantumrelativentropy}.

The dual divergence of $\Div(\rho_1,\rho_2)$ is computed by considering the $(e)$-geodesic connecting $\rho_1$ and $\rho_2$. Let $\rho_1=e^H$, where $H$ is a self-adjoint Hamiltonian. Then,     the $(e)$-geodesic from $\rho_1$ to $\rho_2$ is given by
\begin{equation}
\label{e-geod}
\gamma_e(t)=\frac{e^{H+t\, A}}{\Tr\,e^{H+t\, A}},\quad (t\in[0,1])\,,
\end{equation}
where $A=\log\rho_2-\log\rho_1$ and $e^{H+t\, A}$ denotes the exponential matrix \cite{Petz08}. Since the trace operator is linear in its argument, it commutes with the derivative operator. Therefore, according to Equations               \eqref{m-representation} and                 \eqref{Boge-representation}, we obtain that the $(m)$ and $(e)$ representations of $\dot{\gamma}_e(t)$ are given by
\begin{align}
\label{exponentialm}
& \dot{\gamma}_e^{(m)}=\dot{\gamma}_e(t)=\frac{A\,e^{H+t\,A}}{\Tr e^{H+t\,A}}-\frac{e^{H+t\,A}\,\Tr A e^{H+t\,A}}{\left(\Tr e^{H+t\,A}\right)^2}\,\\
\label{exponentiale}
& \dot{\gamma}_e^{(e)}=\frac{\mathrm{d} }{\mathrm{d} t}\log\gamma_e(t)=A-\frac{\Tr Ae^{H+t\,A}}{\Tr e^{H+t\,A}}\,.
\end{align}

The dual divergence of $\Div(\rho_1,\rho_2)$ is written as follows:
\begin{equation}
\label{dualAyDiv}
\Div^*(\rho_1,\rho_2)=\int_0^1\,t\,\Tr\left(\dot{\gamma}_e^{(m)}\,\dot{\gamma}_e^{(e)}\right)\,d t\,.
\end{equation}

     To perform the computation in Equation               \eqref{dualAyDiv}, we use the expressions of $\dot{\gamma}_e^{(m)}$ and $\dot{\gamma}_e^{(e)}$ given by Equations               \eqref{exponentialm} and                 \eqref{exponentiale}:
\begin{equation*}
\Div^*(\rho_1,\rho_2)=\int_0^1\,t\,\Tr\left(\frac{A^2e^{H+t\,A}}{\Tr e^{H+t\,A}}-2\frac{A\,e^{H+t\,A}\,\Tr A e^{H+t\,A}}{\left(\Tr e^{H+t\,A}\right)^2}+\frac{e^{H+t\,A}\,\left(\Tr A e^{H+t\,A}\right)^2}{\left(\Tr e^{H+t\,A}\right)^3}\right)\,.
\end{equation*}

At this point, we can use the linearity of the trace operator and then the latter expression reduces to:
\begin{equation*}
\Div^*(\rho_1,\rho_2)=\int_0^1\,t \left(\frac{\Tr A^2e^{H+t\,A}}{\Tr e^{H+t\,A}}-\frac{\left(\Tr Ae^{H+t\,A}\right)^2}{\left(\Tr e^{H+t\,A}\right)^2}\right)\,d t=\int_0^1\,t\frac{\mathrm{d}}{\mathrm{d} t}\left(\frac{\Tr Ae^{H+t\,A}}{\Tr e^{H+t\,A}}\right)\,d t\,.
\end{equation*}

Carrying the integration by parts out, we obtain
\begin{eqnarray*}
\Div^*(\rho_1,\rho_2)&=& \left[t\,\frac{\Tr Ae^{H+t\,A}}{\Tr e^{H+t\,A}}\right]_0^1-\int_0^1\frac{\Tr Ae^{H+t\,A}}{\Tr e^{H+t\,A}}\,d t\\
&=& \frac{\Tr \rho_2(\log\rho_2-\log\rho_1)}{\Tr\rho_2}-\left[\log\Tr e^{H+t\,A}\right]_0^1\\
&=&\Tr \rho_2(\log\rho_2-\log\rho_1)- \log\Tr\rho_1\rho_2\rho_1^{-1}+\log\Tr\rho_1\\
&=&\Tr \rho_2(\log\rho_2-\log\rho_1)\,,
\end{eqnarray*}
where we use  $\Tr\rho_1=\Tr\rho_2=1$. This proves that
$$
\Div^*(\rho_1,\rho_2)=\Tr \rho_2(\log\rho_2-\log\rho_1)=\Div(\rho_2,\rho_1)\,.
$$





\section{Conclusions}\label{Section4}

 {As we have demonstrated, for a geometric definition of a general complexity measure, it is important to have a canonical divergence. This paper is based on recent progresses in defining a general canonical divergence within Information {Geometry} \cite{Ay15}, \cite{Felice18}.} This divergence is defined in terms of geodesic integration of the inverse exponential map and holds the geodesic projection property when the structure $(\metric,\nabla,\nabla^*)$ is dually flat \cite{Ay17}. Let $p\in\Ma$ and $\widetilde{\Ma}\subset\Ma$ be a submanifold of $\Ma$, the search for $\hat{p}\in\widetilde{\Ma}$ that minimizes the divergence $\Div(p,q),\,q\in\widetilde{\Ma}$, supplies the solution for defining an information-geometric complexity measure as in Equation               \eqref{complexity}. When every minimizer $\hat{p}$  of the divergence $\Div$ is given by the geodesic projection of $p$ onto $\widetilde{\Ma}$, we say that $\Div$ holds the geodesic projection property. In this regard, the canonical divergence in Equation               \eqref{AyDivnice} would provide a measure of complexity as Equation               \eqref{complexity} for a quite wide range of systems. A further step for defining Equation               \eqref{complexity} for general systems has been put forward in \cite{Felice18}, where  a new divergence  is introduced  that turns out to be a generalization of the canonical divergence in Equation               \eqref{AyDivnice}. As an example of Equation               \eqref{complexity}, we have considered the measure of complexity given by Equation               \eqref{KLcomplexity}, which quantifies how much a probability measure on the product configuration set of the finitely many states on a discrete set $\{1,\ldots,n\}$ deviates from a family of exponential probabilities that amounts to the non-complex set of system states, as it is given by non-interacting states \cite{Aycomplexity}. In this case, the Kullback--Leibler divergence turns out to be suitable for providing the measure of complexity in Equation               \eqref{complexity} for classic states on discrete sets \cite{Ay11}.      {To put the theory of Ay} \cite{Aycomplexity} in perspective and propose the canonical divergence in Equation               \eqref{AyDivnice} as suitable for supplying the complexity in Equation               \eqref{complexity} on general systems, we have then proved that $\Div$ coincides with the {(KL)}-divergence on the simplex of probability measures endowed with the dual structure given by the Fisher metric and the mixture and exponential connections.  

 The quantum counterpart of the general theory yielding the measure of complexity in Equation               \eqref{complexity} does not yet exist. However, a quantum analog of Equation               \eqref{KLcomplexity} has been established on the manifold of positive density operators \cite{Weis15}. Here, the family of non-interacting states is replaced by states  that  are fully described by their restriction to selected subsystems that turn out to be a family of Gibbs states. Therefore, many-party correlations are quantified in the state of composite quantum system, which cannot be observed in subsystems composed of fewer than a given number of parties. The suitable tool for providing such a quantification is established by the quantum relative entropy. This is because the maximum-entropy principle solves the inverse problem to reconstruct a global state from subsystem states and it also gives   a natural scale of many-party correlation in terms of the gap to the maximal entropy value. Hence, the many-party correlation of a quantum state is quantified by the divergence from a family of Gibbs state. The many-party correlation in Equation               \eqref{manyparty} has been implemented in algorithms \cite{Niekamp13} proving to be related to the entanglement of quantum systems as defined in \cite{Vedral97}.      To consider the canonical divergence in      Equation               \eqref{AyDivnice} as an efficient tool for extending the general theory leading to Equation               \eqref{complexity}, we have considered $\Div$ on the manifold of positive density operators with the quantum analog of the Fisher metric and $(m)$, $(e)$ connections induced by the Bogoliubov inner product. We have finally proved that the canonical divergence coincides with the quantum relative entropy.


\vspace{6pt} 







\vspace{6pt} 

\begin{thebibliography}{999}
\bibitem{FeliceChaos}
Felice, D.;  Cafaro, C.;  Mancini, S. Information geometric methods for complexity. {\it Chaos} {\bf 2018}, {\it 28}, 032101.

\bibitem{Aycomplexity}
Ay, N. Information geometry on complexity and stochastic interaction. {\it Entropy} {\bf 2015}, {\it 17}, 2432--2458.

\bibitem{Ay17}
Ay, N.; Jost, J.; Van Le, H.; Schwachh\"ofer, L. {\it Information Geometry,} 1st ed.; Springer International Publishing: Cham, Switzerland, 2017.

\bibitem{Ay11}
Ay, N.; Olbrich, E.; Bertschinger, N.; Jost, J. A geometric approach to complexity. {\it Chaos} {\bf 2011}, {\it 21}, 037103.

\bibitem{Amari00}
Amari, S.-I.; Nagaoka, H. {\it Methods of Information Geometry.} Oxford University Press: Oxford, UK, 2000.

\bibitem{Eguchi85}
Eguchi, S. A differential geometric approach to statistical inference on the basis of contrast
functions. {\it Hiroshima Math. J.} {\bf 1985}, {\it 15}, 341--391.

\bibitem{matumoto1993}
Matumoto, T. Any statistical manifold has a contrast function---on the $C\sp 3$-functions taking the minimum at the diagonal of the product manifold. {\it Hiroshima Math. J.} {\bf 1993}, {\it 23}, 327--337.


\bibitem{AyTusch}
Ay, N.; Tuschmann, W. Duality versus dual flatness in quantum information geometry. {\it J. Math. Phys.} {\bf 2003}, {\it 44}, 1512--1518.

\bibitem{Ay15}
Ay, N.; Amari, S.-I. A Novel Approach to Canonical Divergences within
Information Geometry. {\it Entropy} {\bf 2015}, {\it 17}, 8111--8129.


\bibitem{HJ2}
Felice, D.; Ay, N. Dynamical Systems induced by Canonical Divergence in dually flat manifolds. {\it ArXiv} {\bf 2018}, arXiv:1812.04461.

\bibitem{HJ1}
{Ciaglia, F.; Di Cosmo, F.; Felice, D.; Mancini, S.; Marmo, G.; 
P\'erez-Pardo J.M. Hamilton-Jacobi approach to potential functions in information geometry. {\it J. Math. Phys.} {\bf 2017}, {\it 58}, 063506.}

\bibitem{Felice18}
Felice, D.; Ay, N. Towards a canonical divergence within Information Geometry. {\it ArXiv} {\bf 2018}, arXiv:1806.11363.

\bibitem{Eguchi92}
Eguchi, S. Geometry of minimum contrast. {\it Hiroshima Math. J.} {\bf 1992}, {\it 22}, 631--647.

\bibitem{Weis15}
Weis, S.; Knauf, A.; Ay, N.; Zhao, M.J. Maximizing the divergence from a hierachical model of quantum states. {\it Open Syst. Inf. Dyn.} {\bf 2015}, {\it 22}, 1550006.

\bibitem{Niekamp13}
Niekamp, S.; Galla, T.; Kleinmann, M.; G\"uhne, O. Computing complexity measures for quantum states based on exponential families. {\it J. Phys. A, Math. Theor.} {\bf 2013}, {\it 46}, 125301.

\bibitem{Vedral97}
Vedral, V.; Plenio, M.B.; Rippin, M.A.; Knight, P.L. Quantifying entanglement. {\it Phys. Rev. Lett.} {\bf 1997}, {\it 78}, 2275--2279.

\bibitem{Amari}
Amari, S. {Differential geometry of curved exponential families-curvatures and information loss}. {\it Ann. Statist.} {\bf 1985}, {\it 10}, 357--387.





























\bibitem{Nagaoka95}
Nagaoka, H. Differential Geometrical Aspects of Quantum State Estimation and Relative Entropy. In {\it Quantum Communications and Measurement}; Belavkin V.P., Hirota O., Hudson R.L. Eds.; Springer: Boston, MA, USA, 1995.



\bibitem{Petz08}
Petz, D. {\it Quantum Information Theory and Quantum Statistics;} Springer-Verlag: Berlin/Heidelberg, Germany, 2008.






\end{thebibliography}
\end{document}